\documentclass[lettersize,journal]{IEEEtran}
\usepackage{amsmath,amsfonts}
\usepackage{algorithmic}
\usepackage{algorithm}
\usepackage{array}
\usepackage{tabularx}
\usepackage[caption=false,font=normalsize,labelfont=sf,textfont=sf]{subfig}
\usepackage{textcomp}
\usepackage{stfloats}
\usepackage{url}
\usepackage{verbatim}
\usepackage{graphicx}
\usepackage{cite}
\usepackage{graphicx}
\usepackage{booktabs}
\usepackage{multirow}
\usepackage{threeparttable}
\usepackage{adjustbox}
\usepackage{orcidlink}

\begin{document}

\title{Collaboratively Eliciting Gestures for Geospatial Data Exploration on an MSE with Tangibles and Styluses}

\author{Karen Penaranda Valdivia \,\orcidlink{0000-0002-9883-0816},~\IEEEmembership{Graduate Student Member,~IEEE, \and Nujaimah Ahmed, \,\orcidlink{0009-0004-2938-3890}, \and Aswah Butt \, \orcidlink{0009-0008-2122-1757}, \and \ Mashrufa Orchi \, \orcidlink{0009-0007-0098-5216}, \and  \ Roozbeh Manshaei \, \orcidlink{0000-0001-9336-5831}, \and \ Sarah Hoyos-Hoyos \, \orcidlink{0000-0003-1150-3909}, \and  Emmanuel Kyeremeh \, \orcidlink{0000-0001-8435-3687}, \and  Gabby Resch \, \orcidlink{0000-0002-2842-111X}, \and  \ Robert McLeman \, \orcidlink{0000-0001-9593-1606}, \and \ Jamy Li \, \orcidlink{0000-0003-2440-9719}, ~\IEEEmembership{Member,~IEEE, \and and \ Ali Mazalek \, \orcidlink{0000-0003-0293-5435} , ~\IEEEmembership{Member,~IEEE}}}

\thanks{ \textit{(Corresponding author: Karen Penaranda Valdivia.)} This work involved human subjects in its research. Approval of all ethical and experimental procedures and protocols was granted by the Toronto Metropolitan University (TMU) Research Ethics Board under Application No. 2025-138 and performed in line with TCPS 2 (2022). 

Karen Penaranda Valdivia, Nujaimah Ahmed, Aswah Butt, Mashrufa Orchi, Dr. Roozbeh Manshaei, Dr. Emmanuel Kyeremeh, and Dr. Ali Mazalek are with the Synaesthetic Media Lab, Toronto Metropolitan University, Toronto, ON M5G 1P5, CA (e-mail: karen.penaranda@torontomu.ca).

Sarah Hoyos-Hoyos and Dr. Robert McLeman are with the Department of Geography and Environmental Studies, Wilfrid Laurier University, Waterloo, ON N2L 3C5, Canada.

Dr. Gabby Resch is with the Faculty of Business and Information Technology, Ontario Tech University, Oshawa, ON L1G 0C5, Canada.

Dr. Jamy Li is with the School of Computing, Engineering, \& The Built Environment, Edinburgh Napier University, Edinburgh EH11 4BN, UK. 

The authors were affiliated with the above institutions at the time the work was performed.

This work has been supported by the Social Sciences and Humanities Research Council, the Canada First Research Excellence Fund, the Canada Research Chairs program, the Canada Foundation for Innovation, and the Ontario Ministry of Research and Innovation.

}
\thanks{}}

\markboth{\parbox{6.5in}{\raggedright THIS WORK HAS BEEN SUBMITTED TO THE IEEE FOR POSSIBLE PUBLICATION. COPYRIGHT MAY BE TRANSFERRED WITHOUT NOTICE, AFTER WHICH THIS VERSION MAY NO LONGER BE ACCESSIBLE.}}%
{Shell \MakeLowercase{\textit{et al.}}}
\maketitle

\begin{abstract}
Large tabletop displays and multi-surface environments offer potential for enhancing visual data exploration and collaborative work with geospatial datasets. These systems typically rely on multi-touch interactions, which can pose challenges when the multi-touch sensors misrepresent transitory movements as control inputs, leading to interruptions. Active tangibles and styluses offer an alternative to multi-touch interactions in MSEs, and have shown the potential to facilitate sense-making around large datasets. However, further research is needed to better understand how these modalities can be effectively leveraged for interacting with geospatial data visualizations. To address this, a gesture elicitation study was conducted in which users suggested interactions for 16 geospatial data visualization tasks, presented as a realistic collaborative workflow co-designed with geography and migration researchers. The study produced a taxonomy of user-defined gestures using tangibles and styluses for engaging with geospatial data, along with a thematic analysis of users' experiences with visualization tasks and interaction techniques.
\end{abstract}

\begin{IEEEkeywords}
User centred design, knowledge elicitation, collaboration.
\end{IEEEkeywords}

\section{Introduction}
\IEEEPARstart{M}{ulti-surface} environments (MSEs) and large displays offer considerable benefits for collaborative work in interdisciplinary domains \cite{anastasiou_gesture_2024,wong_collaborative_2013,hwang_exploring_2015,marinho_rodrigues_bancada_2014,tateosian_tangeoms_2010}. MSEs with Tangible User Interfaces (TUIs) have been shown to enhance \IEEEpubidadjcol  users' interactions with data visualizations by facilitating the organization of spatial information \cite{angelini_move_2015,maquil_colortable_104,seyed_eliciting_2012}.  Accordingly, there is a growing interest in the benefits of MSEs for visualizing and interacting with geospatial data and map-based tasks \cite{ens_uplift_2021,nagel_touching_2014,shakeri_hossein_abad_multi_2014,tabrizian_immersive_2016,tateosian_tangeoms_2010}. However, multi-touch has often been used as the primary input modality \cite{marinho_rodrigues_bancada_2014,nagel_touching_2014,shakeri_hossein_abad_multi_2014}, despite prior studies showing that TUIs can outperform it in key tasks for handling geospatial data, such as manipulation and acquisition \cite{al-megren_comparing_2016, tuddenham_graspables_2010}, sorting \cite{al-megren_comparing_2016}, grouping \cite{al-megren_comparing_2016}, and layout manipulation \cite{al-megren_comparing_2016}.  

Multi-touch interfaces also have two-dimensional and technical challenges that can limit users' interactions and interrupt their workflow. TUIs provide a distinct advantage for spatial interactions by offering physical directness that is not bound to two-dimensional surfaces. While multi-touch interfaces allow users to visualize and interact with a large number of objects on a planar surface \cite{schneider_walden_2012},  TUIs support enhanced collaboration, as users can more easily observe and respond to a wider range of physical gestures enabled by the tangible dimension \cite{schneider_walden_2012, brave_tangible_1998}.  Second, multi-touch interfaces may have challenges in supporting users' learning and discovery with data-intensive areas due to finger-size, occlusion, and real estate occupied by Windows, Icons, Menus, and Pointer (WIMP) style controls \cite{drucker_touchviz_2013, vogel_shift_2007, voida_getting_2009, valdes_exploring_2014}.  For example, screen real-estate management could be an issue due to finger-size and occlusion creating barriers for touching small targets \cite{drucker_touchviz_2013, vogel_shift_2007, voida_getting_2009,valdes_exploring_2014}. Additionally, improperly sized WIMP-controls could contribute to cognitive dissonance \cite{drucker_touchviz_2013, valdes_exploring_2014}.  Lastly, technical challenges, such as misrepresenting accidental touches or transient movements as inputs, can lead to workflow interruptions. 

In geospatial contexts, TUIs have primarily been applied to urban planning and collaborative data analysis, often incorporating displays (e.g. \cite{ens_uplift_2021,besancon_hybrid_2017,maquil_geospatial_2015,maquil_towards_2018,maquil_colortable_104, jones_twist_2015, yuan_study_2018, afkari_exploring_2020, sakurai_developing_2021}) , VR or AR systems (e.g., \cite{ens_uplift_2021, maquil_towards_2018}), multi-touch (e.g., \cite{besancon_hybrid_2017,jones_twist_2015, afkari_exploring_2020}), mobile devices (e.g.,\cite{besancon_hybrid_2017}), or malleable materials such as clay or sand (e.g.,  \cite{tateosian_tangeoms_2010, piper_illuminating_2002, tabrizian_immersive_2016}). However,  a gap remains in understanding how tangibles and styluses can be used collaboratively in a map-based context \cite{maquil_towards_2018}, especially since the latter have been shown to support engaging and intuitive user interactions \cite{angelini_move_2015,ens_uplift_2021,jones_twist_2015,manshaei_tangible_2022,valdes_exploring_2014} that could leverage spatial data exploration.  Tangibles have shown promise to support handling of large geospatial datasets \cite{maquil_geospatial_2015}, while supporting distributed and embodied cognition \cite{hornecker_tei_2008}. The precision of styluses has also been found to mitigate many limitations of not having multi-touch \cite{willett_eliciting_2014}, and users can benefit from using styluses for annotations \cite{wobbrock_edgewrite_2003}. As a result, the value proposition of integrating tangibles and styluses could be in their potential to support data-intensive visualizations by leveraging digital inputs and visual feedback through precise hand movements facilitated by the stylus. Further work is needed to understand how tangibles and styluses can be used in a geospatial data exploration context. 

This study aims to investigate how to design interactions for a tabletop MSE that supports collaborative geospatial data exploration using active tangibles and a stylus. An elicitation study was conducted with participants of varying geospatial expertise, who collaboratively co-produced cartographic-perspective and system-perspective tasks while exploring map-based data. Cartographic-perspective tasks are defined as interactions that involve manipulating data within the map itself, while system-perspective tasks refer to interactions involving controls, commands, or menus \cite{shakeri_hossein_abad_multi_2014}. The elicitation study method was chosen as it has proven effective for uncovering user-defined gestures and interaction techniques that might not have been readily apparent to designers \cite{morris_web_2012, leon_eliciting_2024}.  This study and its referents, which are short descriptions of system outcomes that define the expected result of an interaction, were co-designed with migration, geography, and Human Computer-Interaction (HCI) collaborators to reflect realistic workflows and  key geospatial visualization interactions for interdisciplinary migration research teams (e.g., \cite{shakeri_hossein_abad_multi_2014}). 

The contextual need for this work arose from the migration research collaborators' desire for a tool that could allow researchers with varied levels of data literacy and quantitative expertise to collaboratively explore and find patterns and make sense of complex geospatial data. Many of the migration researcher collaborators expressed that their interdisciplinary teams often explored data individually, sharing only completed findings rather than engaging in shared research methodologies. A map-based TUI could help by enabling collaborative map visualizations, which can promote engagement of diverse stakeholders in transparent decision-making \cite{zolnai_map_2014}.  

The research questions addressed by this study are: \\
\textbf{RQ1.} How should tangibles and styluses be employed to manipulate geospatial data visualization in an MSE?\\
\textbf{RQ2.} How should tangibles and styluses be used in a collaborative geospatial data visualization workflow?\\
\textbf{RQ3.} How does participant expertise influence gesture and proposal generation?

To address these questions, 36 participants with varying levels of geospatial data modelling expertise were paired into novice, expert, and mixed pairs and asked to propose interactions for 16 tasks using one or more tangibles, a stylus, or both. It is hoped that the results from this study will inform the future development of a map-based TUI by identifying interaction patterns in how users employ tangibles and styluses to manipulate geospatial data, while enhancing the understanding of their collaborative elicitation process.

Across all expertise levels, participants preferred uni-modal interactions, which aligned with previous findings \cite{leon_eliciting_2024}. The single tangible was the most popular modality and was often used for data exploration and direct manipulation of the mapped layers. The stylus was second and often used for annotations, drawings, or freehand selections. Expanding beyond previous works \cite{leon_eliciting_2024, maquil_towards_2018,valdes_exploring_2014, maquil_geospatial_2015}, it was found that multi-modal inputs (tangible and stylus) complemented each other for compound tasks across tabletop and wall displays; users often confirmed or activated free-hand selections made by the stylus with a tangible. Mixed-expertise pairs produced more matched and simpler proposals than novice or expert pairs, suggesting that cross-expertise dialogue may have influenced creativity and feasibility. This work contributes to design knowledge of user preferences for uni-modal, bi-manual, and multi-modal tangible interactions in collaborative geospatial data exploration, presents consensus analysis and a gesture set of 166 unique interactions, and highlights cross-expertise collaborative interaction patterns.

\section{Related Work} 

This section reviews (1) geospatial tangible user interfaces, (2) collaborative exploration of large datasets with tangibles and MSEs, (3) the elicitation study methodology, and (4) the potential effect of participants' varying expertise. 

\begin{figure}[h]
  \centering
  \includegraphics[width=\linewidth]{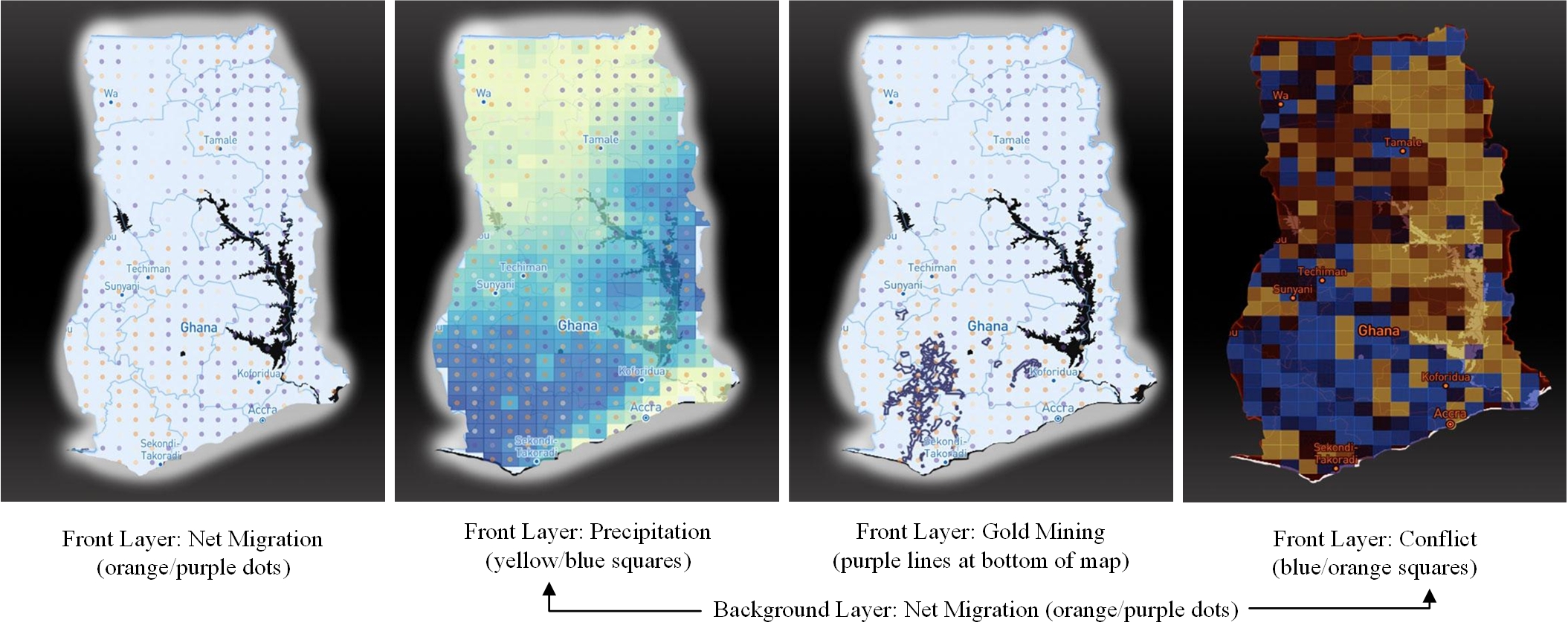}
  \caption{Map layers representing net migration, precipitation, gold mining, and conflict in Ghana. The front layers (precipitation, gold mining, and conflict) are overlaid on the background net migration layer.}
  \label{fig:layers}
\end{figure}

\subsection{Geospatial Tangible User Interfaces }

Advances in GIS have introduced a wide range of features for acquiring, processing, and sharing geospatial information \cite{ goodchild_space_2015}. However, this complexity increases the level of mental effort required for learning and retention, which can detract from knowledge construction \cite{maquil_towards_2018}.  As the field of geographical data visualization evolves from a specialist domain to one involving interdisciplinary actors \cite{hess-luttich_spatial_2011, cay_what_2019}, there is a growing need for geovisualizations that are comprehensible, easy-to-use, and that mitigate cognitive overload \cite{hess-luttich_spatial_2011,cay_what_2019}.  

Maquil et al. \cite{maquil_towards_2018, maquil_geospatial_2015}  introduced a bottom-up approach through \textit{geospatial tangible user interfaces} (GTUI), referring to systems that use physical objects to manipulate large-scale digital maps and geospatial data  projected on a tabletop screen.\footnote{More general work, such as \cite{valdes_exploring_2014}, exploring the design of space gestural interaction with active tokens through user-defined actions, and \cite{manshaei_tangible_2022}, which examines at tangibles for genome data, are also relevant; though not reviewed in detail here based on a focus on geospatial data and combined tangible-stylus exploration.} Their work showed that GTUIs can support face-to-face collaboration and communication, and that playful, motivating tangibles helped multi-expertise users better organize and understand spatial data. For large datasets, tangibles facilitate filtering data \cite{spindler_tangible_2010} and adjusting abstraction levels \cite{elmqvist_hierarchical_2010, spindler_tangible_2010},  both of which are essential for promoting comprehensibility. Styluses can further enhance GTUIs by providing more precise input than fingers, improving the intuitiveness and fluidity of interactions in TUIs \cite{koch_where_2020}. Their annotation capabilities also help compensate for the absence of multi-touch \cite{willett_eliciting_2014}  and some advanced GIS features in GTUIs  \cite{maquil_geospatial_2015}. 

GTUIs have predominantly been used for urban logistics, crisis management, and traffic simulations \cite{maquil_towards_2018}, leaving a gap in understanding how they could support migration studies, another highly interdisciplinary field. At the time of writing this paper, the authors found no prior work that explicitly studied how GTUIs could be incorporated into the context of exploring map-based migration data with both tangibles and styluses.  

\begin{table*}
  \caption{Referent and Task Descriptions}
  \label{tab:referents-perpectives}
  \begin{tabular}{clll}
    \toprule
    \# &Referents & Task Types & Category of Task\\
    \midrule
   1&Navigate map and select Ghana.&  Maneuver/ \cite{brehmer_multi-level_2013,springmeyer_characterization_1992}, Select \cite{brehmer_multi-level_2013, de_morais_bezerra_visual_2014, meyer_four-level_2012, pike_science_2009} & Cartography\\
\hline
    2&Access and display net migration layer.& Accessing extra information \cite{roth_cartographic_2012}, Share \cite{heer_interactive_2012} & Cartography\\
    \hline
    3&Change year of data layer.& Change (range) \cite{brehmer_multi-level_2013, gotz_characterizing_2009} & Cartography\\
    \hline
    4&Add sub-driver layer.& \parbox[t]{4cm}{Add (objects) \cite{brehmer_multi-level_2013}} & System-Perspective\\
    \hline
    5&Identify patterns by annotating.& Annotate \cite{brehmer_multi-level_2013, gotz_characterizing_2009, heer_interactive_2012} & Cartography\\
    \hline
    6&Save visualization.& Save \cite{brehmer_multi-level_2013, liu_mental_2010, roth_cartographic_2012} & System-Perspective\\
    \hline
    7&Move map from table to wall.&  Move\cite{brehmer_multi-level_2013, raskin_humane_2000} & System-Perspective\\
    \hline
    8&Duplicate map on the wall.&  Copy \cite{brehmer_multi-level_2013, raskin_humane_2000} & System-Perspective\\
    \hline
    9&Move map from wall to table.& Move \cite{brehmer_multi-level_2013, raskin_humane_2000} & System-Perspective\\
    \hline
    10&Change driver layer.& Change (range) \cite{brehmer_multi-level_2013, gotz_characterizing_2009} & Cartography\\
    \hline
    11&Create highlight on map.& Highlight \cite{brehmer_multi-level_2013, heer_interactive_2012, raskin_humane_2000} & Cartography\\
    \hline
    12&Brush and link highlight to two maps on wall.& Brushing and linking \cite{roth_cartographic_2012}, Highlight \cite{brehmer_multi-level_2013, heer_interactive_2012, raskin_humane_2000} & Cartography\\
    \hline
    13&Merge maps together.& Merge \cite{brehmer_multi-level_2013, gotz_characterizing_2009} & System-Perspective\\
    \hline
    14&Split.& Split \cite{brehmer_multi-level_2013, gotz_characterizing_2009} & System-Perspective\\
    \hline
    15&Remove, brush, and link.& Delete (objects, sets, graphical objects) \cite{brehmer_multi-level_2013},  Brushing and Linking \cite{roth_cartographic_2012} & System-Perspective\\
    \hline
    16&Delete.& Delete \cite{brehmer_multi-level_2013, roth_cartographic_2012} & System-Perspective\\
    \bottomrule
  \end{tabular}
\end{table*}

Prior research demonstrates the versatility of GTUIs and TUIs across geospatial domains, primarily with non-stylus tangibles. For example, \textit{Mitigation in Urban areas: Solutions for Innovative Cities (MUSIC)} by Maquil et al. \cite{maquil_geospatial_2015}, was a GTUI for urban planning using a tabletop screen with wooden tangible blocks. \textit{Venice Unfolding} by Heidmann, Condotta, and Duval \cite{nagel_venice_2010}  was a GTUI for architectural exploration that used a tabletop screen with polyhedron-shaped tangibles. \textit{Smart City Logistics} by Guerlain, Cortina, and Renault \cite{guerlain_towards_2016} developed a GTUI for exploration of commercial and energy management needs that used a tabletop screen and tangible tokens. \textit{TanGeoMS} by Tateosian et al. \cite{tateosian_tangeoms_2010} was a GTUI for terrain analysis and land management that used map projections onto sand and building-block tangibles. Yuan et al. \cite{yuan_study_2018}  explored industrial and environmental planning with a TUI on a tabletop using IoT blocks, where the former supported dynamic saving of data. \textit{Tangible Globes} by  Satriadi et al. \cite{satriadi_tangible_2022} was a TUI for exploring the tangible-virtual interplay of geospatial data exploration that used AR and tangibles in the form of small globes. \textit{TangibleNet} by Takahira et al. \cite{takahira_tangiblenet_2025} was a TUI for linking map-based nodes on a vertical screen that used projected spatial data and magnet tangibles.   

Despite this breadth, most existing work does not use migration data or explore how a tangible dial and stylus can be used, which may be different than other data or interaction tools.  In addition, many of these past works focus on a single tabletop display and do not explore how tangibles can be used with multiple screens, such as combined tabletop and wall displays (e.g., \cite{maquil_towards_2018,maquil_geospatial_2015, nagel_venice_2010, tateosian_tangeoms_2010}).

\subsection{Collaborative Exploration of Large Datasets with Tangibles and MSEs}

Incorporating MSEs, such as wall and tabletop displays, for big data exploration supports both individual and collaborative visual analytics, promotes equity in physical interactions  \cite{ioannou_tabletops_2016, afkari_exploring_2020},  and enables co-located problem-solving and discussion \cite{isenberg_co-located_2011, afkari_exploring_2020}. The conceptual and trans-disciplinary clarity offered by spatial information can also serve as a powerful enabler in addressing societal challenges \cite{kuhn_core_2012}.  

Extending GTUIs beyond a single tabletop screen to include vertical wall displays (e.g. \cite{maquil_colortable_104,maquil_geospatial_2015, maquil_towards_2018, afkari_exploring_2020, grueau_towards_2016}), offers additional collaborative space interaction space \cite{leon_eliciting_2024},  allows users to analyze and perceive more information at once \cite{pirolli_visual_2001}, and mitigates cognitive overload by helping users to create mental maps with lesser need of virtual navigation \cite{rovine_sketch-map_1989}.   León et al. \cite{leon_eliciting_2024} analyzed how participants explored geospatial data using an interactive wall display, including styluses but not tangible pucks, and found that it facilitated exploration. Moreover, a review of 28 papers on multi-surface interactions specifically with geospatial data found that the most used interaction devices were handheld devices (e.g., phones, tablets), computers, and wall displays \cite{shakeri_hossein_abad_multi_2014}. These papers suggest the importance of wall displays in geospatial data exploration, despite their exclusion from some past works. 

\subsection{Elicitation Studies}
 
Elicitation studies are an interaction design methodology focused on identifying user-defined interactions \cite{wobbrock_user-defined_2009}.  In these studies, participants are presented with system outcomes, known as \textit{referents}, and then asked to propose gestures, known as \textit{symbols}, that would trigger those outcomes \cite{wobbrock_user-defined_2009,leon_eliciting_2024}.  Researchers group these symbols into unique sets of gestures, called \textit{signs}, and derive a \textit{consensus set}, capturing the most agreed-upon interaction for each referent \cite{wobbrock_user-defined_2009,leon_eliciting_2024}. The elicitation study methodology was appropriate for this study since GTUIs employing tangibles and stylus-based interactions on MSEs remain largely under-explored, and the methodology enables the discovery of users' interaction techniques, tendencies, and preferences that might not have been readily apparent in the initial design phase \cite{morris_web_2012}.

The elicitation study methodology is also suitable for uncovering consensus in users' proposed multi-modal interactions \cite{leon_eliciting_2024}, such as with this study's combined use of tangible(s) + stylus. Some notable examples of elicitation studies relevant to the present work include Valdes et al. \cite{valdes_exploring_2014} and León et al. \cite{leon_eliciting_2024}.  Valdes et al. \cite{valdes_exploring_2014} had 19 users do a query-building task for big data (i.e., genomic application) on a large display (i.e.,  tabletop screen) with tangible and gestural language inputs. They identified a user-generated gesture vocabulary of 21 distinct gestures. León et al. \cite{leon_eliciting_2024} had 20 users do an interaction elicitation study in pairs to explore how users used touch, speech, pen, and mid-air gesture manipulations while exploring data collaboratively on a large display (i.e., vertical screen). They identified 1015 proposals, with participants often favouring touch and speech modalities.  

Elicitation studies can also probe how expertise shapes interaction proposals, which is important in domains like migration research, where familiarity with geospatial modelling varies. While several elicitation studies have recruited participants based on expertise \cite{morris_web_2012, leon_eliciting_2024} or prior interface experience \cite{leon_eliciting_2024}, the authors do not know of any past elicitation studies that explicitly compared participants' expertise and their performance. This gap may be surprising considering the importance of literature on expertise in design \cite{cross_expertise_2004}. 

\section{Study Design}

This section covers the apparatus and input modalities, the participants, the referents, and the study procedure. 

\subsection{Apparatus and Input Modalities}

The study took place in a university lab equipped with 55-inch MultiTaction Cell screens (16:9 aspect ratio). The wall configuration consisted of nine vertical screens, arranged as five extended displays, while the tabletop setup included three screens arranged as two extended displays. Two chairs were placed in front of the tabletop, providing users the flexibility to sit, stand, or walk around the tabletop and vertical screens during the study. To support documentation of the interactions for data analysis, the study included an observer and two recording cameras on opposite ends of the tabletop.

\begin{figure}[h]
  \centering
  \includegraphics[width=0.6\linewidth, height=40 mm]{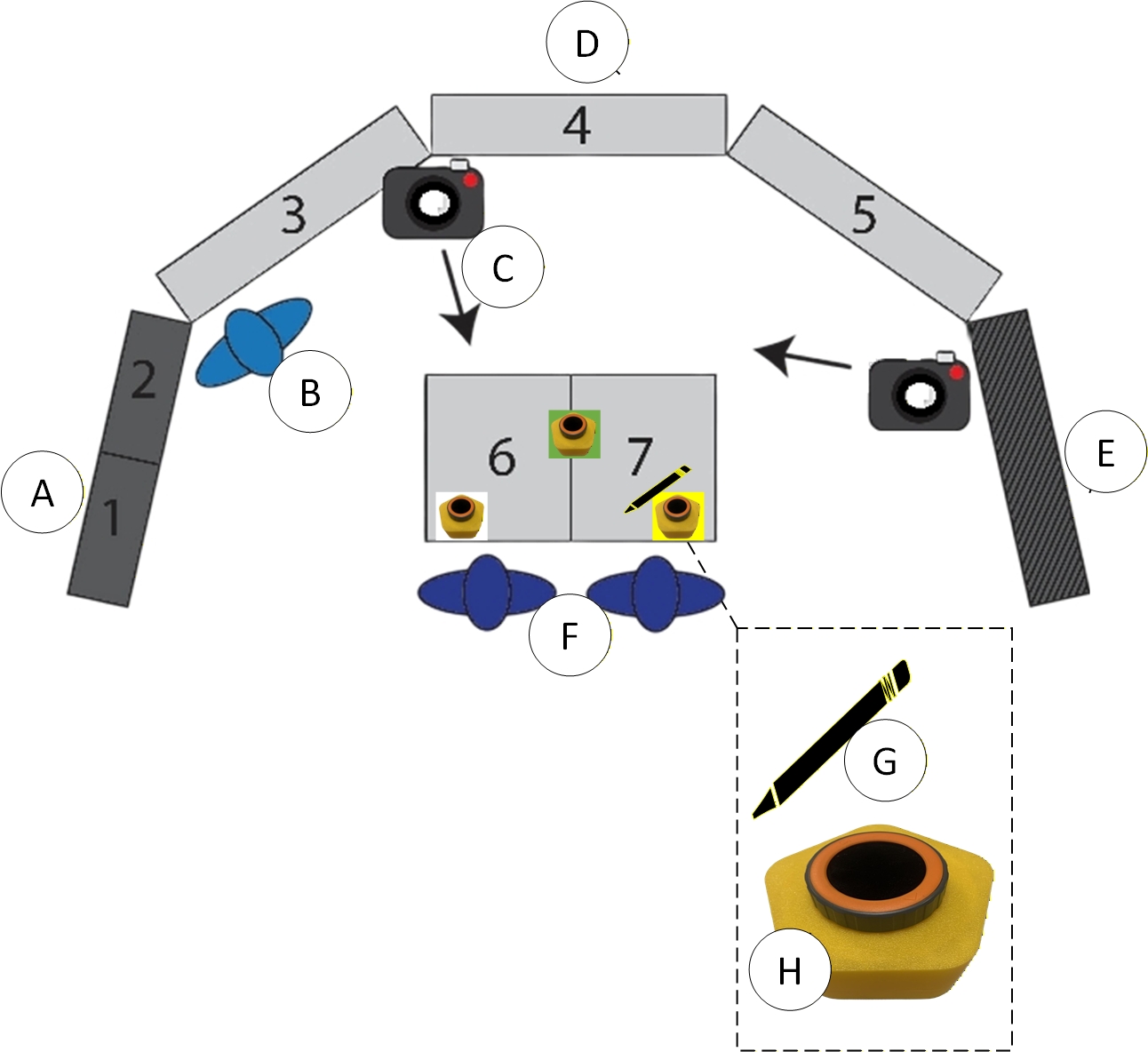}
  \caption{The set-up includes: A) referent displays (screens 1-2); B) 1 observer; C) 2 cameras for documentation; D) MSE composed of vertical wall displays showing visualizations (screens 3-5) and tabletop display for interactions (screens 6-7); E) unused display screen; F) participants A and B; G) 1 stylus; and H) 3 tangibles. Participants could sit, stand, or walk-around.}
  \label{fig:set-up}
\end{figure}

Each pair received pentagon-shaped tangible dials (representing active tangibles) and a pen (representing a stylus). Each tangible featured an M5Stack Dial, ESP32-S3 Smart Rotary Knob with a 1.28-inch round touchscreen, encased in a 3-inch, 3D-printed pentagonal case (see Fig.\ref{fig:set-up}). The dial was selected for its rotary encoder for menu selections, flexible USB-C or battery power options, and development flexibility, which facilitates integration with the JavaScript-based Mapbox interface. The tangible's pentagon case was inspired by five major drivers of migration: environmental, socio-cultural, economic, demographic, and political \cite{toronto_metropolitan_university_what_2024}. This shape enabled the design team to explore how physical affordances might influence user gestures. However, the study's facilitators did not enforce any expectations, giving participants full autonomy to decide whether or not to incorporate the tangible's shape into their interactions. The tangible cases were printed in three arbitrary colours: yellow, green, and white.

The stylus was chosen to support natural and expressive inputs, including annotating, drawing, tapping, and circling or lassoing \cite{gero_eliciting_2022}. Its precision could mitigate the limitations introduced by the lack of multi-touch input \cite{willett_eliciting_2014}. Multi-touch on the tabletop and wall displays was intentionally disabled to encourage collaboration and prevent workflow disruptions. During initial design and testing of map-based interactions on the tabletop display, some users' transitory movements were mistakenly treated as commands. Compared to multi-touch, TUIs can support enhanced collaboration by enabling users to have a greater range of physical gestures that are not bound by the 2D dimension  \cite{schneider_walden_2012, brave_tangible_1998}. Multi-touch may disrupt collaborative elicitations due to the influence of finger sizes, occlusion, and the size of WIMP-controls \cite{drucker_touchviz_2013, vogel_shift_2007, voida_getting_2009, valdes_exploring_2014}.  

\subsection{Participants}

Thirty-six participants with self-described novice or expert geospatial modelling knowledge were recruited, and grouped into 18 pairs: 6 expert-expert, 6 mixed, and 6 novice-novice. Most participants (72.2\%) were around 20 to 30 years old, and 60.0\% were undergraduate or graduate university students. Educational backgrounds spanned social sciences, media studies, environmental and aeronautical engineering, and geographical sciences. Of the 36 participants, 11\% reported using MSEs daily, 20\% reported using MSEs 2 to 3 times per week, while 66.0\% indicated they rarely or never used MSEs. Despite this, 61.0\% held a positive perception of MSEs prior to the study, driven by either curiosity or prior experience, while 35.0\% were indifferent or uninterested. In terms of stylus use, only 11.0\% reported using a stylus at least once per week or daily, while 57.0\% responded "never" or "rarely." Among those who did use a stylus, 26.0\% reported using it for digital drawing, and 23.0\% for note-taking or annotation. Notably, all stylus users used them in conjunction with handheld devices. 

\begin{table}[htbp]
\caption{Definitions and Illustrations Used to Code Participants' Proposed Gestures}
\label{tab:Definitions}
\footnotesize
\setlength{\extrarowheight}{0pt}
\renewcommand{\arraystretch}{0.85}
\begin{tabularx}{\columnwidth}{@{}l X@{}}
\toprule
Gesture & Definition Used for Coding \\
\midrule
Place \includegraphics[width=1cm,height=0.5cm]{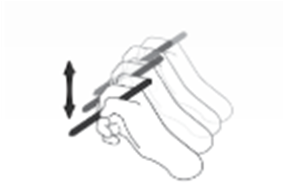} 
& Changing position through the air without contacting a surface in the middle of the movement action. \\
Press \includegraphics[width=1.5cm,height=0.5cm]{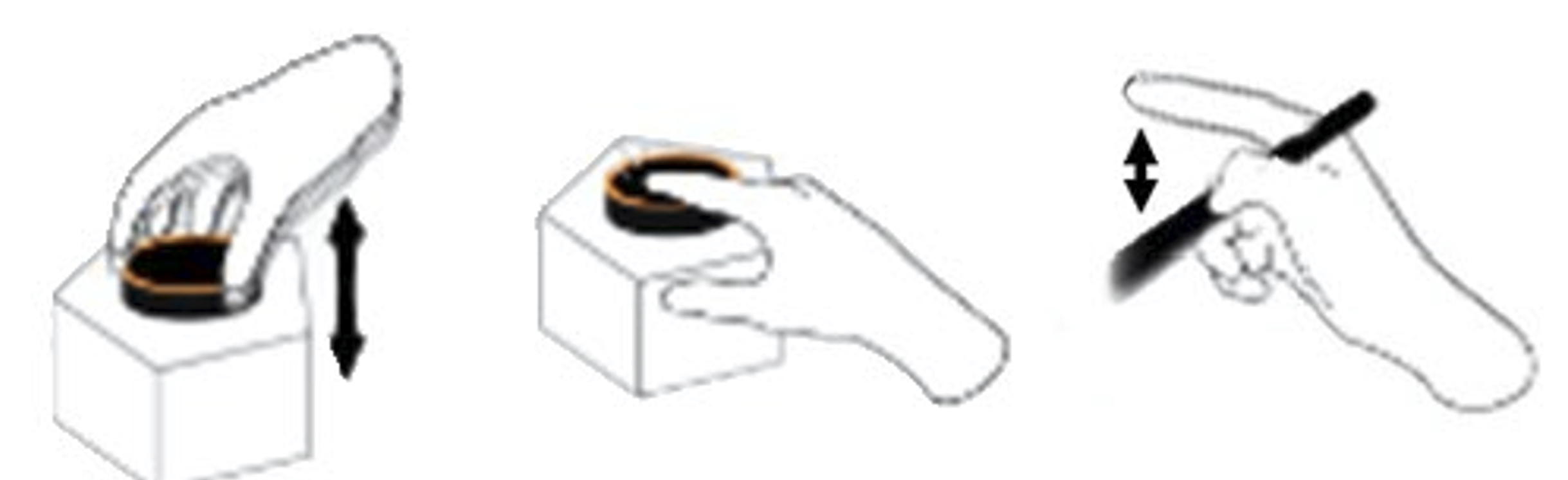} 
& Tap with the finger, touch with the finger, applying pressure with finger(s) onto object, except for swipe. \\
Slide \includegraphics[width=1.25cm,height= 0.5 cm]{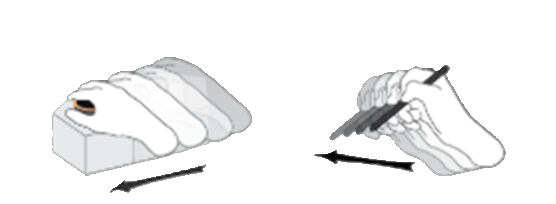} 
& Tangibles keep surface contact, move the tangible(s) or stylus across the screen(s). \\
Rotate \includegraphics[width=1cm,height=0.5 cm]{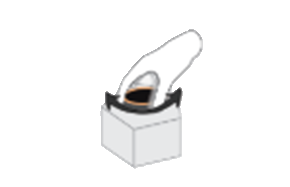}  
& Turn or pivot around a fixed point while holding a part (i.e., dial) or entire body of input devices (i.e., pentagon body or stylus). \\
Lasso  \includegraphics[width=1cm,height=0.5cm]{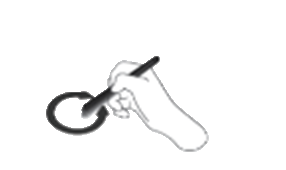} 
& Make a selection by creating a circular or loop-shaped path. The selection must be enclosed. Ensure surface contact with screen(s). \\
Draw \includegraphics[width=1cm,height=0.5cm]{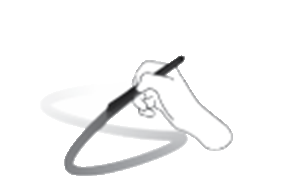} 
& Make lines or shapes by moving across the screens. The selection does not have to be enclosed. \\
\bottomrule
\end{tabularx}
\end{table}

\subsection{Referents}

The elicitation study was co-designed by a team of migration researchers, geographers, data scientists, and HCI researchers with varying levels of geospatial modelling expertise.  The team curated a realistic workflow of 16 task-oriented referents for map-based visualizations or traditional GUI tools (refer to Table \ref{tab:referents-perpectives}). 
Referents' visualizations were created from Mapbox-based screenshots of data layers for net-migration, precipitation, gold-mining, and conflicts in Ghana from 1985 to 2010 (see Fig \ref{fig:layers}), constrained by available coded data for that region and period. The referents covered both cartographic-perspective tasks, involving direct manipulation of map data \cite{shakeri_hossein_abad_multi_2014}, and system-perspective tasks, involving interaction with menus and controls \cite{shakeri_hossein_abad_multi_2014}. 

Since a realistic geospatial workflow also involves repetitive interactions, such as loading data for multiple layers, this study aimed to reduce task-redundancy via some referents that required each participant to achieve the same result but with different time periods (i.e., referents 2 or 3) or data-layer selections (i.e., referents 2, 4, 6, 7, and 8). This motivated the use of two screens (1 and 2) (see Fig.~\ref {fig:set-up}) to present referent descriptions and visualizations, and for selected referents 2-4, 6-8, to display alternate descriptions or corresponding map views. 

\subsection{Elicitation Study Procedure}

This study's procedures aimed to mitigate the legacy biases of elicitation studies \cite{besevli_investigating_2018,leon_eliciting_2024, morris_web_2012} by employing \textquotesingle priming\textquotesingle{}, \textquotesingle production\textquotesingle{} and \textquotesingle partners\textquotesingle{} \cite{morris_web_2012} in their activities. Before starting, each pair received three coloured tangible dials (active tangibles) and a pen (stylus). They were asked to complete four activities: (1) view an introductory (priming) presentation to briefly review the elicitation study methodology and geospatial data exploration; (2) complete a pre-task survey to assess their geospatial knowledge; (3) participate in the elicitation study in partners, such that production was encouraged through referents that promoted their collaborative design of open-ended tasks (proposals) \cite{danielescu_iterative_2022}; and (4) complete a post-task questionnaire to reflect on their experience. 

To mimic a realistic workflow of handling geospatial data, each pair (participants \textit{A} and \textit{B}) were asked to role-play as migration researchers. For each of the 16 referents, \textit{A} and \textit{B} were shown descriptions and their corresponding map visualization(s) on wall displays 1 and 2 (shown in Fig. \ref{fig:set-up}). Then, \textit{A} and \textit{B} were given two minutes to verbally brainstorm proposal interactions. Next, \textit{A} and \textit{B} wrote and optionally sketched two proposals each on paper, with \textit{A}  focusing on wall display 2 and \textit{B} on wall display 1, and with the requirement that at least one proposal per participant incorporates a tangible. After, \textit{A} and \textit{B} were asked to physically perform and verbally describe their four gesture proposals to each other. 

Three coders then independently reviewed the interaction proposals, corrected or supplemented the proposal details based on video recordings and optional participant drawings, and then standardized following a procedure similar to that outlined by Morris \cite{morris_web_2012} and León \cite{leon_eliciting_2024}. This standardization involved categorizing proposals into unique signs based on the input modalities used, the data attributes that were manipulated, and the end goal or outcome of the interaction.

\section{Results}
This section presents results for the maximum consensus (MC), consensus-distinct ratio (CDR), and statistical analysis. 
 
\subsection{MC and CDR Analysis}
Coders identified 1152 proposals, which were grouped into 166 unique interactions or \textit{signs} (based on distinct proposals per referent). The gesture coding framework, including key definitions and illustrative examples, is summarized in Table \ref{tab:Definitions}. 

Since standard agreement scores for elicitation studies do not account for multiple proposals per referent \cite{leon_eliciting_2024, vatavu_clarifying_2022}, this study instead uses Morris' consensus metrics (i.e., MC and CDR) \cite{leon_eliciting_2024, morris_reducing_2014, vatavu_clarifying_2022, morris_web_2012} to capture the frequency of the top proposals when each participant contributed two proposals per referent. These metrics, as shown in Table ~\ref{tab:interaction_analysis} were chosen as high MC and CDR values may be indicative of strong agreement on a primary interaction with few close competitors \cite{morris_web_2012}. To illustrate, higher MC values may indicate that the proposed interactions are more intuitive among participants\cite{danielescu_iterative_2022}, and higher CDR values may indicate that there is a greater proportion of interactions that exhibit some consensus among participants \cite{vatavu_between-subjects_2016}. 

Across all referents and modalities, the mean MC was 18.6\% , and using a consensus threshold of 2, the mean CDR was 0.212. Among participants, interactions with \textquotesingle 1 tangible only\textquotesingle{} tended to exhibit similar commonalities ($\overline{MC}_{1 \ tangible \ 
only}= 14.9\%$) as interactions with \textquotesingle 1 stylus only\textquotesingle{}  ($\overline{MC}_{1\ stylus\ only}= 14.7\%$). Interactions with  \textquotesingle 1 stylus only\textquotesingle{} may show more intuitive or natural usage patterns, as the mean CDR of these interactions is higher than that for  \textquotesingle 1 tangible only\textquotesingle{}   ($\overline{CDR}_{1 \ stylus \ only}= 27.5\%$ \ versus\  $\overline{CDR}_{1 \ tangible \ only}=23.7\%$).  

Multi-modal interactions had much lower mean MC and CDR values compared to the means from the interactions with \textquotesingle 1 tangible only \textquotesingle{} or \textquotesingle 1 stylus only \textquotesingle{}.  The mean CDR values were: $\overline{MC}_{tangible(s) \ + \ stylus} = 1.2\%$,  $\overline{CDR}_{tangible(s) \ + \ stylus} = 7.4\%$, $\overline{MC}_{tangibles \ ( \geq 2 )} = 1.6\%$,  and $\overline{CDR}_{tangibles only \ ( \geq 2 )} = 5.7\%$.  In comparison with uni-modal interactions, the low mean MC and CDR values may indicate that multi-modal interactions had more individualization and diverse user behaviours than those with single-input tools. The complexity of the users' proposals may have also been higher with multi-input tools, but it is important to note that the number of multi-elicitations is much less than uni-modal interactions, as shown in Table \ref{tab:Breakdown}. 

\begin{table}
\centering
\caption{Breakdown of Signs by Modality and Expertise}
\label{tab:Breakdown}
\begin{tabular}{l c l}
\hline
Modality & \# Signs & Contribution by Expertise\\
\hline
1 Tangible Only & 93 & E: 64, M: 68, N: 57 \\
1 Stylus Only & 63 & E: 41, M: 47, N: 33 \\
Tangible(s) + Stylus & 4 & E: 4, M: 1, N: 2 \\
Tangibles Only $(\geq 2)$ & 6 & E: 3, M: 5, N: 2 \\
\hline
\end{tabular}
  \begin{tablenotes}
  \scriptsize
    \item\(^{a}\) E = expert pair, M = mixed pair, N = novice pair. 
    \item\(^{b}\) E, M, N pairs may have generated the same unique proposals.
  \end{tablenotes}
\end{table}

\subsection{Statistical Analysis}
 
Two sets of statistical tests were conducted: one treating each proposal as the unit of analysis and another treating each participant as the unit. For proposal-level analyses, Pearson's Chi-Squared tests were used for categorical measures (e.g., proposal matching, modality). Analysis of Variance (ANOVA) was used for interval measures (e.g., proposal length) when comparing the three types of pairs (expert, mixed, novice). For participant-level measures, Welch two-sample t-tests were used to compare the two knowledge levels (expert vs. novice). All statistical tests were performed using R (version 4.4.1) and RStudio (version 2024.04.2+764).

\begin{table*}[t]
  \caption{Maximum Consensus (MC) and Consensus Distinct Ratio (CDR) by Modality for User-Elicited Interactions}
\label{tab:interaction_analysis}
  \centering
  \scriptsize
  \begin{tabular}{c l l c c c c c c c c c c}
    \hline
    \# & Referent & MCI(s) & \multicolumn{2}{c}{All I.M.} & \multicolumn{2}{c}{I.M. 1} & \multicolumn{2}{c}{I.M. 2} & \multicolumn{2}{c}{I.M. 3} & \multicolumn{2}{c}{I.M. 4} \\
    & & & MC & CDR & MC & CDR & MC & CDR & MC & CDR & MC & CDR \\
    \hline
    1 & Navigate map, select Ghana & [Slide, Rotate, Press] & 14\% & 0.204 & 14\% & 0.208 & 6\% & 0.235 & -- & -- & 6\% & 0.333 \\
    2 & Access net migration layer & [Rotate, Press] & 25\% & 0.191 & 25\% & 0.227 & 11\% & 0.188 & -- & -- & 6\% & 0.167 \\
    3 & Change data layer year & [Press, Rotate] & 22\% & 0.208 & 22\% & 0.240 & 11\% & 0.267 & -- & -- & -- & -- \\
    4 & Add sub-driver layer & [Rotate, Press, Press] & 14\% & 0.304 & 14\% & 0.435 & 11\% & 0.400 & -- & -- & -- & -- \\
    5 & Annotate patterns & [Lasso] & 31\% & 0.118 & 8\% & 0.430 & 31\% & 0.182 & 6\% & 0.200 & -- & -- \\
    6 & Save visualization & [Press] & 28\% & 0.248 & 28\% & 0.250 & 22\% & 0.286 & 3\% & 0.330 & -- & -- \\
    7 & Move map table to wall & [Slide] & 22\% & 0.229 & 14\% & 0.269 & 22\% & 0.231 & -- & -- & 6\% & 0.250 \\
    8 & Duplicate map on wall & [Press, Slide] & 11\% & 0.148 & 11\% & 0.143 & 8\% & 0.250 & -- & -- & -- & -- \\
    9 & Move map wall to table & [Slide] & 11\% & 0.208 & 11\% & 0.250 & 11\% & 0.222 & -- & -- & -- & -- \\
    10 & Change driver layer & [Press, Rotate] & 22\% & 0.286 & 22\% & 0.267 & 8\% & 0.600 & -- & -- & -- & -- \\
    11 & Create map highlight & [Lasso] & 11\% & 0.208 & 11\% & 0.250 & 11\% & 0.222 & -- & -- & -- & -- \\
    12 & Brush link 2 wall maps & [Press, Slide] & 11\% & 0.131 & 11\% & 0.214 & 6\% & 0.087 & -- & -- & -- & -- \\
    13 & Merge maps & [Slide] & 17\% & 0.136 & 6\% & 0.250 & 17\% & 0.111 & -- & NA & -- & -- \\
    14 & Split maps & [Rotate, Press] & 11\% & 0.269 & 11\% & 0.261 & 8\% & 0.385 & -- & -- & 8\% & 0.167 \\
    15 & Remove/brush/link & [Place, Place] & 22\% & 0.261 & 17\% & 0.280 & 22\% & 0.286 & 6\% & 0.333 & -- & -- \\
    16 & Delete & [Draw] & 19\% & 0.250 & 14\% & 0.304 & 19\% & 0.308 & -- & -- & -- & -- \\
    \hline
  \end{tabular}
  \vspace{-0.25mm}
  \begin{tablenotes}
  \scriptsize
    \item\(^{a}\) MCI = Most common interaction(s); I.M. = Input modality; I.M. 1 = tangible only; I.M. 2 = stylus only; I.M. 3 = tangible + stylus; I.M. 4 = tangibles only $(\geq 2)$.
    \item\(^{b}\) "--" = \# Unique Proposals is 1 (less than threshold of 2) so NULL. 
    \item\(^{c}\) NA = No Proposals elicited.
  \end{tablenotes}
\end{table*}

\subsubsection{Effect of Pair Type}
Proposal-level analysis (i.e, whether the given proposal was repeated by another participant for a given referent by pair type)  using Pearson's Chi-squared test was significant  $\chi^{2}(2) = 15.5 , \quad p < 0.001$. Proposals from individuals in mixed pairs were more likely to be matched by another participant than those from expert or novice pairs. The count of non-matched proposals generated by individuals in mixed versus expert versus novice pairs is 200 > 168 > 146 proposals, and comparatively for non-matched proposals is 184 < 216 < 238. This suggests that mixed pairs generated higher-quality (more likely to be common) proposals than those in uniform pairs.

An ANOVA on proposal length (i.e., number of steps in a proposal) with pair type as a between-proposal variable was also significant $F(2, 1149) = 32, \quad p < 0.001, \quad \eta_{p}^{2} = 0.05$. Proposals from individuals in mixed pairs had the shortest number of steps, followed by proposals generated from those in expert pairs and then those generated in novice pairs, $M = 2.65$ , $SD = 1.13$ vs $M = 2.82$, $SD = 1.15$ and $M = 3.39$, $SD = 1.66$.

A Chi-squared test on modality by pair type was not significant, $\chi^{2}(6) = 10.4 , \quad p = 0.011$. 

Treating the unit of analysis as an individual participant, an ANOVA on the total number of matched proposals out of 32 with pair type as a between-participant variable was marginally significant, $F(2, 33) = 3.1$, $p = 0.57$, $\eta_{p}^2 = 0.16$. Participants in mixed pairs produced a higher number of matched proposals, $M = 16.7$, $SD = 4.1$, vs. expert pairs $M = 14$, $SD = 3.5$ and vs. novice pairs $M = 12.2$, $SD = 5.5$, though it did not reach significance.

An ANOVA test on the average length of proposals with pair type as a between-participant variable was significant, $F(2,33) = 6.0$, $p = 0.006$, $\eta_{p}^2 = 0.27$. Participants in mixed pairs produced shorter interactions than those in uniform pairs (especially versus novice pairs), $M = 2.65$, $SD = 0.36$, vs. expert pairs $M = 2.82$, $SD = 0.51$, and novice pairs $M = 3.39$, $SD = 0.72$. Follow-up Chi-squared tests on each motion type (e.g., Slide, Rotate, etc.) found that mixed pairs used fewer Lasso motions and that novice pairs used more Rotate, Press and Place motions $p_s \leq 0.013$. Therefore, participants in mixed pairs performed better than those in uniform pairs.

A Chi-squared test on the average number of proposals designed per modality by user knowledge was not significant, $\chi^{(2)}(6) = 0.86$ , $p = 0.99$.

\subsubsection{Effect of Expertise}
Treating the unit of analysis as a proposal, Pearson's Chi-squared test on the proposal modality (e.g., "1 Tangible Only") by user knowledge (i.e., "Expert" or "Novice") was not significant, $\chi^{(2)}(3)= 5.5$, $p = 0.14$. Pearson's Chi-squared test on the proposal matching (i.e., whether the given proposal was repeated by another participant for a given referent) by user knowledge was also not significant, $\chi^{(2)}(1)=1.5$, $p = 0.21$. A Welch two-sample t-test on the proposal length with user knowledge as a between-proposal variable was significant, $t(1087) = -3.8$, $p < 0.001$. Proposals produced by experts were simpler (shorter) than those made by novices, $M = 2.8$, $SD = 1.2$ vs $M = 3.1$, $SD = 1.5$ . Follow-up Chi-squared tests on each motion type (e.g., Slide, Rotate, etc.) found that novices used more Place motions than experts, $\chi^{(2)}(1)=10$, $p <0.01$.

Treating the unit of analysis as a participant, a Welch two-sample t-test on the total number of matched proposals out of 32 comparing experts with novices was not significant, $t(32) = 0.78$, $p = 0.44$. A Welch two-sample t-test on the mean length of proposal comparing experts with novices was also not significant, $t(28) = -1.5$, $p = 0.14$. A Chi-squared test on the average number of proposals designed per modality by user knowledge was not significant,$\chi^{(2)}(3) = 0.30$, $p = 0.96$ (see Table ~\ref{tab:Breakdown}).

\section{Discussion}

\subsubsection{Input Modality Preferences for Cartographic or System-Perspective Tasks}

The  \textquotesingle 1 tangible only\textquotesingle{}  had the highest MC values across 9 out of the 16 referents (i.e., referents 1, 2, 3, 4, 8, 9, 10, 12, 14  in Table \ref{tab:interaction_analysis}).   This was followed by the \textquotesingle 1 stylus only \textquotesingle{}, which had the highest MC values across 7 out of the 16 referents (i.e., referents 5, 6, 7, 9, 11, 13, 15, 16 in Table \ref{tab:interaction_analysis}).  There was a tie with referent 9, and the full breakdown of the interactions with the highest MC values by task-type is shown in Table \ref{tab:distribution_MCs}.  The \textquotesingle 1 tangible only \textquotesingle{} and the \textquotesingle 1 stylus only \textquotesingle{} had higher user-preferences and agreements compared to the \textquotesingle tangible(s) + stylus \textquotesingle{} or \textquotesingle tangibles only \begin{math} (\geq 2) \end{math} \textquotesingle{}, but the tangible was favoured for cartographic tasks, and the stylus was favoured for more system-perspective tasks. 

\subsubsection{Proposals by Knowledge Level}

Participants in mixed expertise teams performed better (i.e., more matched and simpler proposals) than those in uniform expertise teams, especially versus novice teams. No evidence was found that the proposal modalities used differed across pair type or user knowledge groups, with a single tangible or a single stylus preferred in most cases.

\subsubsection{Participants Monitored Each Other's Activity Within Personal Spaces and Often Provided Feedback}

Participants frequently looked over each other's work while documenting their elicited proposals. Some participants voiced their feedback to add additional detail or modifications.  Differences emerged, however, between expert, mixed, and novice pairs.  

Three out of six expert pairs were particularly vocal about how gestures should be documented, often interrupting each other with "what if" questions, and over-focusing on potential failure modes. This occasionally led to territorial behaviour when one partner altered the other's proposal. Some experts drifted into suggesting their ideas for new referents or functionalities, which were documented for future work, but required the study's facilitators to refocus them on the current task to finish the study. 

Mixed pairs often exhibited more passive exchanges than expert groups. However, two out of six mixed groups had experts who voiced some concern with how the novices provided feedback of the proposals. In one pair, a novice repeatedly interrupted the expert while seeking reassurance, which the expert later described as irritating and wished the novice had shown more confidence. In another pair, the expert had to repeatedly encourage a timid novice to speak up, highlighting how knowledge imbalances can create hesitancy and skew proposals toward one participant’s ideas.

Novice participants tended to be more talkative with each other and less concerned about how their partners documented their elicitations. Five out of six novice pairs exhibited fluid and supportive communication when one of their partners requested documentation support. During these conversations, however, sometimes novice members added additional gestures to their initial proposals. These seemed less constrained by legacy biases or physical restrictions in comparison to the suggestions made by expert or mixed pairs. For example, one novice pair imagined throwing the tangible in the air from the tabletop to the wall-display to project a map.  Another pair discussed the possibility of throwing the tangible in the air and letting it fall onto the tabletop to delete the maps. Expert and mixed pairs mostly focused on planar gestures with few in-air actions (or collisions). 

\subsubsection{Supporting an Embodied Metaphor of  Migration Drivers through the Tangible's Pentagon Shape}
The tangible used in this study was a low-fidelity prototype with a pentagonal case, embodying a metaphor of the five macro-drivers of migration (i.e, environmental, sociocultural, economic, demographic, and political \cite{toronto_metropolitan_university_what_2024}). Most participants did not immediately recognize this metaphor, but after the priming presentation, some began to prefer using the tangible for performing cartographic interactions and the stylus for system-perspective interactions (refer to Table \ref{tab:distribution_MCs}). To illustrate, participants tended to grab a tangible to manipulate map layers from the tabletop to the wall displays, and they also favoured using different coloured tangibles to represent distinct layers. 

\subsubsection{Potential Benefits of the Tangible(s) and Stylus for the Final System}
The tangible(s) and the stylus supported interactive communication and collaboration across users of mixed geospatial expertise and professions. Brainstorming with both input modalities on the tabletop created a safe shared workspace for novices and experts to exchange ideas and see beyond legacy biases. Indeed, this shared workspace was not free from the tensions of participants' disagreements, mimicking real-life scenarios that collaborative migration or policy researchers face. However, the tangible(s) and the stylus afforded users a visual means to convey gestures that might not have been fully apparent by verbal communication. In turn, this supported conflict-resolution by balancing novices' creativity with experts' insights on feasibility and utility. Employing the tangible(s) and the stylus in the final system may support migration and policy researchers in their understanding of how diverse users process and interact with spatial information; an interdisciplinary understanding of the latter can become an enabler for solving social issues \cite{kuhn_core_2012}.

\setlength{\tabcolsep}{2pt} 

\begin{table}[htbp]
  \caption{Input Modality of Interactions with Highest MC per Referent}
  \label{tab:distribution_MCs}
  \centering
  \scriptsize
  \begin{tabular}{@{}p{3.2cm}ccc@{}}
    \toprule
    Task Type & 1 Tangible & 1 Stylus & Tie \\
    \midrule
    Cartography        & 5 & 1 & 0 \\
    System Perspective & 3 & 6 & 1 \\
    \bottomrule
  \end{tabular}
\end{table}

\section{Limitations} 

Recruitment predominantly drew on a migration institute at an undergraduate-focused university, with over 90\% of participants under 40. Participants were paired by self-reported knowledge level and scheduling rather than by prior familiarity. To better capture the diversity of migration researchers, future work will include delocalizing the research with a focus on broadening age groups and pairing familiarity. Further study is also required to understand how diverse users' metaphorical mappings would change with the addition of other input modalities in a working prototype. Nevertheless, the employment of diverse knowledge groups to collaboratively propose cartographic and system-perspective tasks with tangibles highlights the potential of this study to inform the design of intuitive spatial data-exploration.  

\section{Conclusion}
This elicitation study showed how expert and novice users can propose cartographic and system-based interactions for a collaborative geospatial visualization. The pentagon-shaped tangible, embodying the five macro-drivers of migration \cite{toronto_metropolitan_university_what_2024},  worked well for directly manipulating map layers, while the stylus was better suited to perform annotations, drawings, free-hand selections, and dragging or moving tangible-based selections. Together, the tangible-only and the stylus-only modalities complemented one another. 

Despite differing geospatial expertise, participants converged on a preference for uni-modal interactions, with tangible-only the most frequently used modality and stylus-only a close second (refer to Table \ref{tab:interaction_analysis}).  Multi-modal modalities (i.e., \begin{math} \geq 2 \end{math} tangibles or the tangible(s) + the stylus) were used less often for completion of complex tasks, perhaps influenced by this study's 16 referents that offered a low-level workflow with limited compound tasks. Future work will address this limitation through a working prototype that will present more complex modelling requiring higher-level tasks. 

Across all referents and modalities, both experts and novices showed generally low MC and CDR values (refer to Table \ref{tab:interaction_analysis}), with differences between groups not reaching statistical significance, suggesting similar perceptions of task demands and complexity. Future work will broaden the participant pool in terms of age and discipline and use a task-based study with a working prototype to ensure accessibility. It is hoped that the findings of this study will extend beyond migration studies, informing the design of collaborative tangible systems that support mixed-methods researchers in discussion and evidence-based decision-making.

\bibliographystyle{IEEEtran}
\bibliography{references}

\newpage

\vfill

\end{document}